\documentclass[a4paper,12pt,notitlepage,twoside]{article}
\usepackage{amsmath}
\usepackage{hyperref}
\usepackage{epsfig}
\usepackage{comment}
\begin{document}

\centerline{\bf THE MEAN-FIELD DYNAMO MODEL}

\begin{center}{M. Yu. Reshetnyak \\ \vskip 0.5cm  {\small  \it 
 Schmidt Institute of Physics \\ of the Earth of the Russian Academy of Sciences 
Moscow, Russia \\
{\bf m.reshetnyak@gmail.com} }
}
\end{center}

{\small \noindent The  2D Parker's mean-field dynamo equations with a various distributions of the $\alpha$- and $\omega$-effects are considered. We show that smooth profiles of $\alpha$ and $\omega$ can produce dipole configuration of the magnetic field with the realistic magnetic energy spectrum. We emphasize that fluctuating $\alpha$-effect leads to increase of the  magnetic energy  at the small scales, breaking the dipole configuration of the field. The considered geostrophic profiles of $\alpha$ and $\omega$ correspond to the small-scale polarwards/equatorwards  travelling  waves with the small dipole field contribution. The same result is observed for the dynamic form of the $\alpha$-quenching, where two branches of the weak and strong solution coexist. 
 \vskip 0.1cm}


\section{Introduction}
\label{section:1}

The last decades demonstrated impressive success of the three-dimensional geodynamo modelling. It appears that combination of the compositional and thermal convection can drive the dynamo mechanism, transforming the heat and kinetic energies to the energy of the magnetic field. To the moment, the various 3D geodynamo models can reproduce the main features of the observable geomagnetic field: the dipole structure, reversals and excursions of the field, as well as the regimes without reversals, which correspond to the chrons, well known in palaeomagnetism \cite{RK13}.

However analysis of the data simulated in the 3D models sometimes is not easier task rather the analysis of the observations itself. Moreover, due to specific of the 3D modelling, which requires the detailed resolution of the small-scaled turbulence, it appears impossible to reproduce the long-time sequences of the magnetic field evolution, compared with the paleo- and archemagnetic  observations. Note that from the point of view of the observer, information that can be derived from the bulk of 3D data in the models, is excessive, because it 
 can not be verified by the observations with the pure resolution. 

The tendency of the geodynamo development only proves this statement: to reach the desired parameter regimes of the magnetohydrodynamic (MHD)  process in the turbulent liquid core one needs  at least $10^{24}$ grid points for the data simulations  that corresponds to the Reynolds number ${\rm Re}=10^9$. In the same moment, the direct observations of the geomagnetic field, limited by the screening of the low-conductive mantle, are bounded with the first decade of the spherical functions (more precisely $n\le 13$)  \cite{IGRF}.
It means that scale to scale comparison of 3D simulations with observations is only possible in the negligible part of the spatial scales, involved into the simulations. On the other hand,  the time scales in 3D models are order of magnitude shorter than the geological times.

This discrepancy  results in renovation of the quite old mean field approach in geodynamo, which is able to reproduce behaviour of the large-scale magnetic field. In its turn,  simulated  large-scale magnetic field  already   can be easily compared with the  observations. Due to reduction of the 3D basic equations to the axi-symmetric form, this approach permits to simulate long-term evolution of the magnetic field, compared with the palaeomagnetic records.

The mean-field theory  was developed by the two independent scientific groups. The exhaustive theoretical background was elaborated in the German group \cite{KR},  mostly concentrated on the astrophysical applications. The main  result of the theory is the description of the large-scale magnetic field generation with the conductive turbulent me\-dium and velocity shear. The back-reaction (or quenching) of the magnetic field onto the flow was introduced by the damping of the turbulence.

 The other, geophysical approach, developed by the Russian scientist S. I. Braginsky, included influence of the magnetic field onto  the large-scale velocity field. The famous geodynamo Z-model  could reproduce the dipole structure of the magnetic field and made a 
 remarkable insight into the physics of the liquid core \cite{Br75}. One of the crucial points of  this model is existence of the strong magnetic wind, which corresponds to the large magnetic field counterpart to the azimuthal force in the Navier-Stokes equation. That was the reason of the too strong toroidal magnetic field, compared to the poloidal part.

Only latter it was recognised  that influence of the magnetic field on the flow is a very delicate process 
 \cite{BS05},
 \cite{HR10}:  magnetic field does not change the cyclonic form of the flow essentially, as well as it does not produce too large azimuthal velocity
\cite{Jones}. This is the motivation to consider the classical mean-field dynamo  equations without back-reaction of the magnetic field on the large-scale flow, using only $\alpha$-quenching, concerned with the damping of the turbulence by the large-scale magnetic field. So far the kinetic energy of the turbulence is smaller than that of the large-scale velocity,  suppression of the turbulence looks more  acceptable. 

This approach is supported by the new knowledge on  the hydrodynamic of the liquid core: the spatial distribution of the differential rotation and kinetic helicity in the rotating spherical shell, where the geostrophic state holds 
\cite{R10}. The other point is the study of  the more complex quenching mechanisms of the  $\alpha$-effect  \cite{KRR}, developed after the first success of the mean-field theory, and its influence on the magnetic dipole behaviour. 

 We also consider applications of the  popular approach of the fluctuating $\alpha$-effect \cite{Hoyng}, and discuss  constraints  on the amplitude of such  fluctuations, which follows from the form of the spatial spectrum of the  geomagnetic field \cite{Langel}.

\section{Basic Equations and Methods of Solution}
\label{section:2}
The mean magnetic field $\bf B$  is governed by the induction equation
\begin{equation}\label{1}
{\partial{\bf
B}\over\partial t}=\nabla \times \Big( \alpha\,{\bf B}+
{\bf V}\times {\bf B}
-\eta\, {\rm rot}{\bf B}   \Big), 
\end{equation}
where $\bf V$ is the large-scale velocity field,  $\alpha$ is the $\alpha$-effect,  and $\eta$ is a magnetic diffusion.

 The  magnetic field ${\bf B}=\left( {\bf B^p},\, {\bf B^t} \right)$
has two parts: the poloidal component ${\bf B^p}=\nabla\times {\bf A}$, and the toroidal component $\bf B^t$, where $\bf A$ is the  vector  potential of the magnetic field.
 
The principal  point of the mean-field dynamo theory is the separation of the physical 
 fields onto the large- and small-scale counterparts. Information on the large-scale velocity field is described by $\bf V$, and on the small-scale fields fluctuations by the $\alpha$-effect.

 Usually, it is supposed that the mean field $\bf B$ has axial symmetry.
 This assumption follows from the effect of the differential rotation, which suppresses deviations of the frozen magnetic field into the flow   from the axial symmetry.

Due to the axial symmetry of the magnetic field, vector potential $\bf A$ and $\bf B^t$ have 
the only one azimuthal component
 in the spherical system of coordinates
$(r,\,\theta,\, \varphi)$: ${\bf A}(r,\,\theta)=(0,\, 0,\, A)$, and ${\bf B^t}(r,\,\theta)=(0,\, 0,\, B)$. Then the poloidal field  can be written in the form: 
\begin{equation}\label{2}
\displaystyle
{\bf B^p}=
\left(
 {1\over r\, \sin\theta}{\partial\over\partial \theta }\left( A\, \sin\theta \right),\,
-{1\over r} {\partial \over \partial r}  \left( r\, A \right),\, 0
 \right).
\end{equation}

In terms of scalars $A$ and $B$ Eq(\ref{1}) is reduced to the following system of equations:
\begin{equation}\label{3}
\begin{array}{l}
\displaystyle
{\partial{
A}\over\partial t}=\alpha {B} + \left({\bf V}\times\,{\bf B}\right)_\varphi
+ \left( \nabla^2 - {1\over r^2\sin^2\theta}  \right) {A}
\\  \\ 
\displaystyle
{\partial{B}\over\partial t}={\rm rot}_\varphi \left( \alpha\,{\bf B} +
 {\bf V}\times\,{\bf B}\right)+
\left( \nabla^2 -{1\over r^2\sin^2\theta}  \right){B},
\end{array}
\end{equation}
where  the subscript $\varphi$ corresponds to the azimuthal component of the vector.

Eqs(\ref{3}),   solved in the spherical shell
$r_i\le r\le r_\circ$ with $r_i=0.35$, $r_\circ=1$,
 are closed with the pseudo-vacuum boundary conditions: ${ B}=0$, and $\displaystyle {\partial \over\partial r} \left( r A\right)=0$ at $r_i$ and $r_\circ$
 and $A=B=0$ at the axis of rotation $\theta=0,\, \pi$. The simplified form of the vacuum boundary condition  for $A$ is  well adopted in dynamo community, and presents a good approximation of the boundary with the non-conductive medium 
\cite{Jouve}. The reason why
 the vacuum boundary condition is  used at the inner core boundary is discussed in 
 \cite{R13} and concerned with the weak influence of the inner core on the reversals statistics of the magnetic field \cite{Wicht}.

In the general case velocity  $\bf V$ is a three-dimensional vector, as a function of $r$ and $\theta$. Further we consider only the effect of the differential rotation, concerned with the $\varphi$ component of $\bf V$, leaving the input of the meridional circulation  $(V_r,\,V_\theta)$ out of the scope of the paper.

For the quite large amplitudes 
 of  $\alpha$ and $\bf V$ solution $(A,\, B)$ grows exponentially, and one needs to introduce
 the feedback of the magnetic field
 onto the sources of the input energy $\alpha$ and $\bf V$. 
 As we already mentioned above, we concentrate our study on the feedback of the magnetic field onto the $\alpha$-effect, responsible on the production of the large-scale poloidal magnetic fields by the small-scaled turbulence. This approach let us to bypass  solution of the Navier-Stokes equation, which, at least in the geodynamo, is the most difficult part of the full dynamo problem. We recall that
turbulent  convection presents  at the small scales, where the magnetic field is already absent due to the high magnetic diffusion. The ratio of the diffusion scales of the  velocity and magnetic fields is of the order of the Roberts number ${\rm q}=10^{-5}$, which is quite small in the liquid core. However the magnetic field is not generated at the small-scaled part of the kinetic energy spectrum, to get a self-consistent solution for the velocity field, one needs to solve the Navier-Stokes equation in the full range of scales. This task is still out of reach of the modern computer facilities.

 Here we specify two forms 
 of the $\alpha$-quenching. The first one, the so-called algebraic quenching, originates from the simple idea of the damping of the $\alpha$-effect's amplitude with the mean magnetic field:
\begin{equation}\label{4}\displaystyle
\alpha={\alpha_\circ(r,\,\theta)   \over 1+\displaystyle {E_m(r,\,\theta)\over E_m^\circ}},
\end{equation}
where $E_m=B^2/2$ is the magnetic energy, and  $E_m^\circ$ is the  constant parameter. The choice of this parameter relates 
 to our 
 assumptions on the ratio of the kinetic to magnetic energies, and depends strongly on the angular rotation of the body
\cite{RS03}. 

The more sophisticated form of the $\alpha$-quenching follows from 
 \cite{PFL}, 
\cite{ZRS},  where influence of the magnetic field onto the $\alpha$-effect was described  by  the magnetic pat of the $\alpha$-effect, so that the total effect is the sum: $\alpha=\alpha_h+\alpha_m$. Here  $\alpha_h$ and $\alpha_m$ are the hydrodynamic (the so-called kinetic $\alpha$-effect) and magnetic parts, correspondingly. The damping of $\alpha$ means generation of $\alpha_m$ with the opposite sign to  $\alpha_h$. This idea was  formulated latter in the form of the evolutionary equation for $\alpha_m$ \cite{KRR}, see for details Section 6.
  This kind of the $\alpha$-quenching, derived from the basic MHD equations, leads to oscillations in the system, and was used in the solar dynamo to mimic the solar cycle of the magnetic activity.

The differential operators in Eqs.(\ref{3}--\ref{4}) were   
 approximated with the second-order central-differences   scheme in space, and  integrated in 
 time using the second-order Runge-Kutta method. These algorithms resulted in  C++ object oriented code with use of Blitz++ C++ library for the easier compact operations with the arrays. The  post-processor graphic visualization was organized using the Python graphic library MatPlotlib. All simulations were done under the Linux OS.

The code passed the set of the benchmarks. The first  one is  the free-decay mode test for the diffusion operator in the equation: 
\begin{equation}\label{5}\displaystyle
{\partial A\over \partial t}=\left( \nabla^2 - {1\over r^2\sin^2\theta}  \right) {A}, 
 \end{equation}
with $A=0$ at the axis and at $r=r_i$, and 
 $\displaystyle{\partial\over \partial r} \left( r\, A\right)=0$ 
 at $r_\circ$.
 
Simple analytic solution of (\ref{5}) for  testing can be written in the form: 
\begin{equation}\label{6}\displaystyle
  A= e^{\gamma\, t}\, \left(j_1(\sqrt{\lambda}\,r)+C\, y_1(\sqrt{\lambda}\, r)\right)P_1^1,
 \end{equation}
 where 
 $P_1^1$ is the associated Legendre polynomial of degree $1$, and order $1$, and $j_1$, $y_1$ are the spherical Bessel functions of the first and second kind.
 Note, that in contrast to the  scalar Laplace equation, where  the axially symmetric meridional part of the solution is described by  $P_l^0$, the order of our vector diffusion operator's eigenfunctions is shifted by one, and corresponds to  $P_1^1$.

 Putting expression for $A$ (\ref{6}) in (\ref{5}), and 
using boundary conditions for $A$ at the radial boundaries, 
 one has 
condition of solvability for $\lambda$. Solution of this transcendent equation, using package of the analytic algebra SymPy for the Python, gives $\lambda=4.8732823108648490873$, 
 that leads to  $C=0.380157168844938$ and $\gamma=-23.74888048138824458988$. This estimate of 
 $\lambda$, $C$, and $\gamma$  is enough to satisfy to the  boundary conditions with the double precision accuracy, used in the program. Using  (\ref{6}) as the initial condition, we simulated 
Eq.(\ref{5}) and obtained the decay rate equal to the analytic $\gamma$ with the accuracy up to 0.5\% 
 for  $N_r\times N_\theta$ mesh grid points, with $N_r=N_\theta=101$.

The other test was  the benchmark on the threshold of the magnetic field generation, the Case $A'$ from 
\cite{Jouve}, which was also passed successfully.
 
\section{Simple Forms of $\alpha$-$\omega$ Profiles}
\label{section:3}
We start from the simple forms of  the $\alpha$-effect and azimuthal velocity $V_\varphi$, adopted  in the mean-field dynamo.

 From the general arguments it is known that the $\alpha$-effect has the dipole symmetry in respect  to the equator plane. We also assume that it is positive in the northern hemisphere, so that 
\begin{equation}\label{7}\displaystyle
\alpha_\circ=C_\alpha\sin\left(\pi{r-r_i\over r_\circ-r_i}\right)\sin(2\theta),
\end{equation}
 where $C_\alpha$ is a positive constant. This assumption is in agreement with that fact that kinetic helicity $\chi$
  is negative in the northern hemisphere and $\alpha\sim - \chi$    \cite{KR}. In  (\ref{7}) $\alpha$ vanishes at the poles, which, as we see below, is also the good approximation of the real $\alpha$-effect,  derived from 3D models    \cite{R10}.

For the azimuthal velocity we take 
\begin{equation}\label{8}\displaystyle
V_\varphi=C_\omega(r-r_i)(r_\circ-r)e^{\displaystyle-0.7^{-1}\left(\theta-{\pi\over 2}\right)^2} \sin(\theta),
\end{equation}
 where $C_\omega$ is the amplitude. This profile is  symmetric to the equator plane and has maximum at the equator for $C_\omega>0$.

As follows from analysis of 1D Parker's equations, which can be derived from Eqs(\ref{3}), neglecting  $r$-derivatives, solution depends on the product of ${\cal D}=C_\alpha\, C_\omega$, called the dynamo number. Change of the sign of $\cal D$ leads to the change of direction of propagation of the dynamo wave. In general, in 2D case this statement is not correct, and the direction of the wave propagation depends on the spatial distribution of $\alpha$ and $V_\varphi$. It means that choice of signs of $C_\alpha$, $C_\omega$ needs additional information.

\begin{figure}[t!]
\def\ss{.55}
\def\ssize{9.2cm}
\vskip -4.0cm
\hskip 1.0cm 
\begin{minipage}[t]{\ss\linewidth}
\hskip -1.95cm 
\includegraphics[width=\ssize]{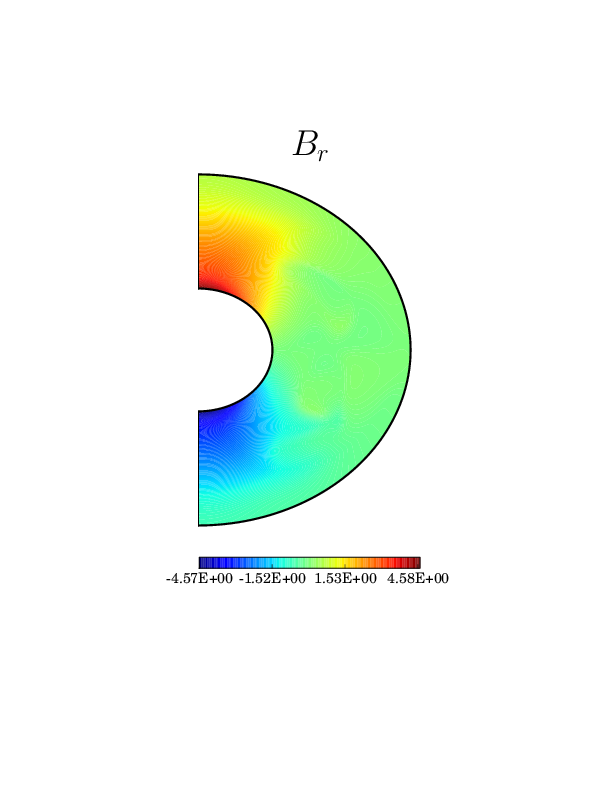}
\end{minipage}\hfill
\hskip -15.cm 
\begin{minipage}[t]{\ss\linewidth}
\includegraphics[width=\ssize]{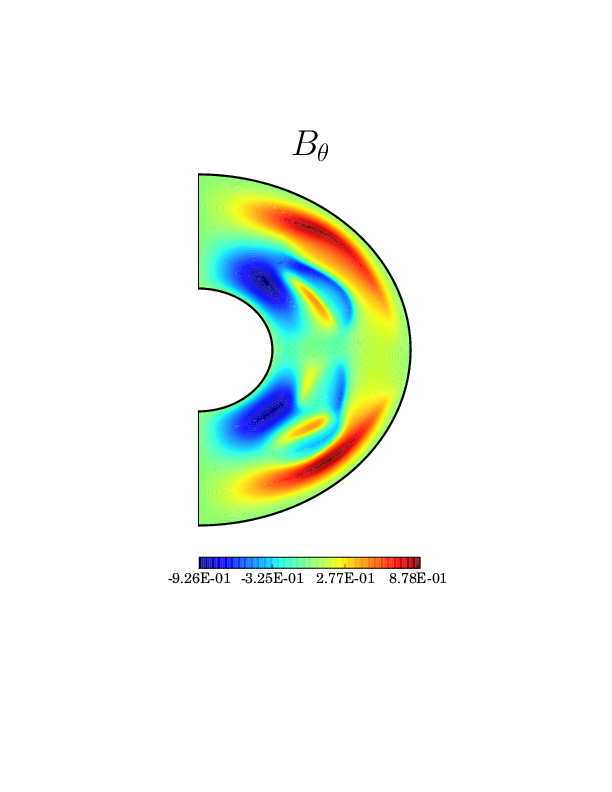}
\end{minipage}
\vskip -4.5cm
\hskip -1.0cm
\begin{minipage}[t]{\ss\linewidth}
\includegraphics[width=\ssize]{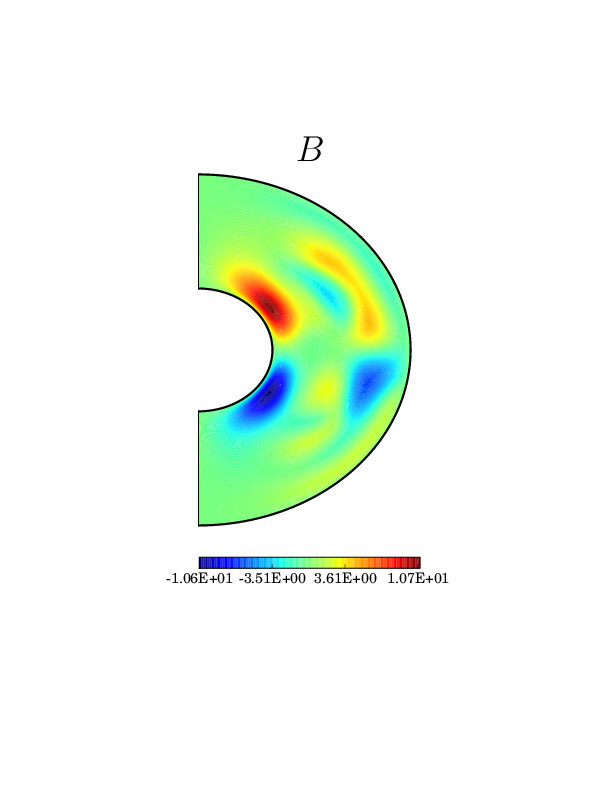}
\end{minipage}\hfill
\hskip -1.6cm
\begin{minipage}[t]{\ss\linewidth}
\includegraphics[width=\ssize]{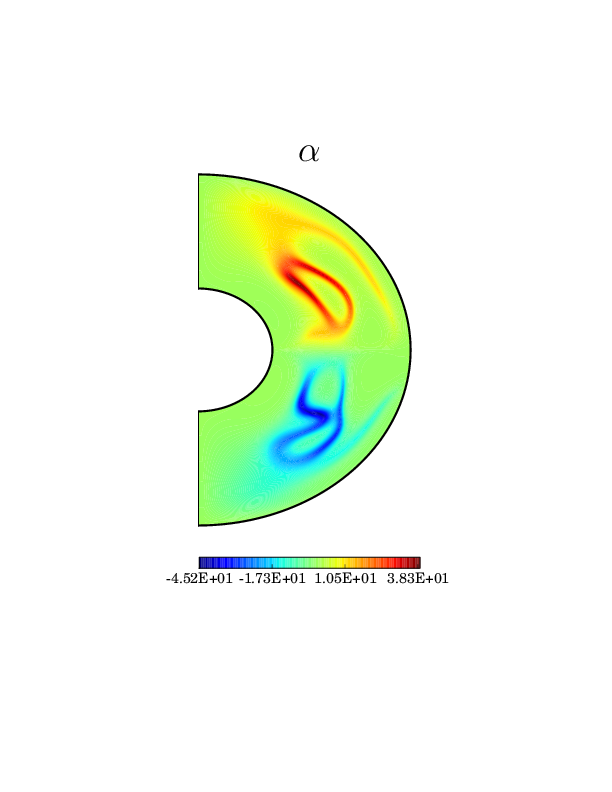}
\end{minipage}
\vskip -2.5cm
 \caption{Spatial distribution of
 $B_r$, $B_\theta$, $B$, and $\alpha$ for $C_\alpha=50$, $C_\omega=5\, 10^4$. 
} \label{fig1}
\end{figure}

The  positive sign of $\alpha$ follows from the  simple relation $\alpha\sim -\chi$ between the $\alpha$-effect  and kinetic helicity $\chi$, which is negative in the northern hemisphere. From 3D geodynamo simulations follows that $V_\varphi$  has maximum in the bulk of the liquid core at the equator plane 
 \cite{R10}, what is also is in agreement with the  helioseismological observations in the solar convective zone  \cite{Bel}. These two arguments fix the signs of $C_\alpha$ and $C_\omega$. 

Integration in time of Eqs(\ref{3}--\ref{4}) with $\alpha_\circ$, and $V_\varphi$, given by (\ref{7}--\ref{8}), with the time step $\tau=10^{-6}$, leads to the  quasi-periodic  oscillatory solution, which has the dipole symmetry for $B_r$ and $B$, and the quadrupole type for $B_\theta$, see Fig.~\ref{fig1}. Note that 	magnetic field is mostly concentrated inside of the spherical shell, in spite on the penetrating poloidal component of the field outside of the shell.
 Solution is highly non-linear, what  is proved by the very irregular distribution of the $\alpha$-effect, damped with the magnetic field, see Eq.(\ref{4}). 

Evolution of the axi-symmetric magnetic dipole $g_1^0$, which contributes to the 
 axi-symmetric form of the 
Mauersberger-Lowes spectrum \cite{Langel} 
$S_l=(l+1)\, \left(g_l^0\right)^2,$
 corresponds to the regime in oscillations, where  the mean level of the field is larger rather the amplitude of its fluctuations. The range of oscillations is  $(0.4-0.47)$, and the dipole does not   reverse. This regime is the typical example of the $\alpha\omega$-dynamo with the poloidal magnetic energy $\displaystyle{1\over 2}\left(B_r^2+B_\theta^2\right)$ of factor 30 smaller than the toroidal one, $\displaystyle B^2/2$. The ratio of the dipole to quadrupole components $S_1/S_2\sim 10$ is quite large. and remains large for $C_\alpha=(5- 500)$ for the fixed value of $C_\omega$.

\begin{figure}[ht!]
\def\ssize{11cm}
\vskip -5.5cm
\hskip 4.15cm 
\includegraphics[width=\ssize]{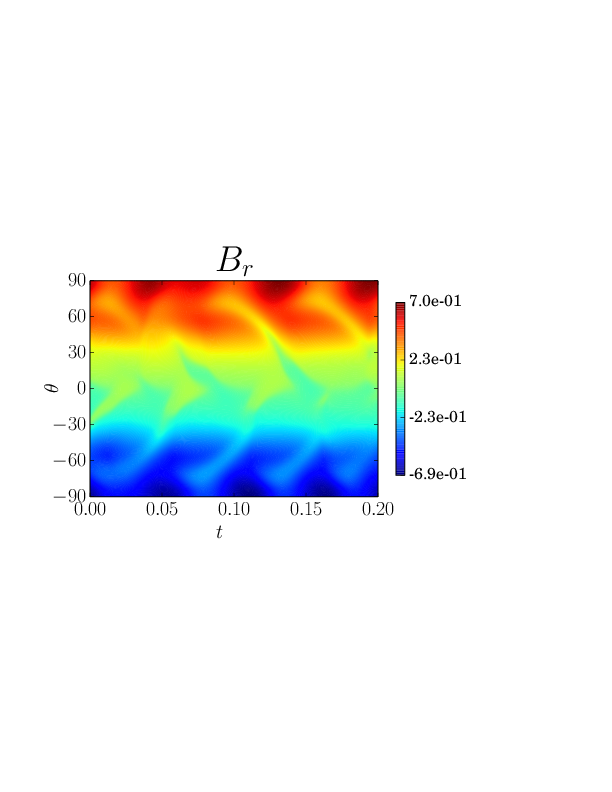}
\vskip -8.7 cm
\hskip 4.15cm \includegraphics[width=\ssize]{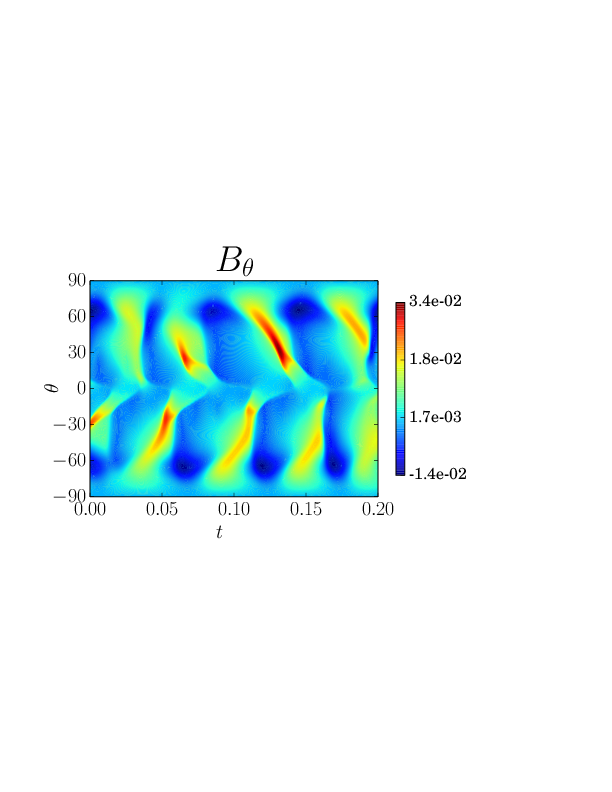}
\vskip -8.7 cm
\hskip 4.15cm \includegraphics[width=\ssize]{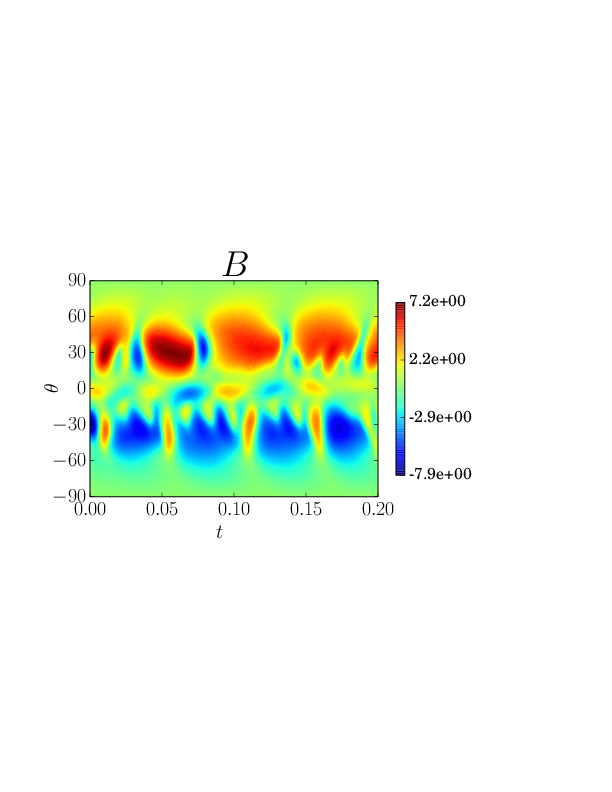}
\vskip -4.5cm
 \caption{The butterfly diagrams for 
 $B_r$, $B_\theta$ at $r=r_\circ$, and $B$ at $r=0.7$.
} \label{fig2}
\end{figure}

To follow the details of the magnetic field generation we consider the butterfly diagrams of the magnetic field, Fig.~\ref{fig2}.
 The poloidal field $(B_r,\, B_\theta)$ is taken at the outer boundary, and the toroidal one at the maximum of generation, near the inner boundary. The poloidal field demonstrates two kinds of the waves, propagating to the equator at $|\theta|<80^\circ$, and
 to the poles  at $|\theta|>70^\circ$. Note that there is intersection of the  waves in the band $\theta=70-80^\circ$. Simultaneous   existence of the   polarwards and equatorwards waves is the subject of debates in the solar dynamo 
\cite{Moss11}. These waves can be related to the quasi-periodic archeomagnetic waves, which also demonstrate different directions of  propagation.

 The toroidal magnetic field $B$ near the inner boundary oscillates  at the  non-zero mean level, and  at least potentially can contribute to the torsional oscillations, concerned with the inner-outer cores interaction.  The absolute maximal values of the azimuthal field $B$ in the northern  hemisphere is  shifted relative to the field in the southern hemisphere  at the half of the period of oscillation. It means that  solution can not be described with the combination of a few   symmetric and antisymmetric functions relative to the equator plane,  and that it has more complex structure.

\section{Random $\alpha$ }
\label{section:4}

The proposed axi-symmetric $\alpha\omega$-model is a crude simplification of the original 3D MHD equations at least in that sense that $\alpha_\circ$, which describes production of the magnetic field with the turbulence,  is a constant parameter. In the more consequent approach \cite{Hoyng} $\alpha_\circ$ has a random fluctuating part, caused with the finite number of the fluid cells. This assumption leads to the reasonable estimates of the $\alpha$ fluctuations in the solar convective zone  \cite{Moss}. This approach was used to get a spontaneous reversals of the magnetic field in the finite-dimensional geodynamo model
 \cite{Sobko}.  

However, we have to use results of the finite-dimensional geodynamo models  very carefully because 
 the considered Galerkin decomposition in \cite{Sobko} included only two first modes. On the other hand, input of the energy by the fluctuating   $\alpha$ at the small scale can change the magnetic field  spectrum   essentially. So far there is no  inverse cascade in the $\alpha\omega$-equations, as it happens, e.g., in 2D hydrodynamic turbulence   \cite{KM80},
  energy of fluctuations will not transfer over the spectrum to the large scales, and concentrate at the scale of fluctuations. It can happen that such energy injection will lead to the change of the spectrum. In its turn, increase of the energy at the small scales will result in disagreement with observations, which demonstrate predominance of the magnetic dipole component on the higher harmonics. We recall that as it follows from the practice of the 3D dynamo simulations,  solution is  decently resolved if the kinetic and magnetic energies drop by more than a factor of 100 from
the spectral maximum to the cut-off wavelength \cite{C99}.


\begin{figure}[ht!]
\def\ss{8.2cm}
\vskip -3.5cm
\hskip 3.5cm 
\includegraphics[width=\ss]{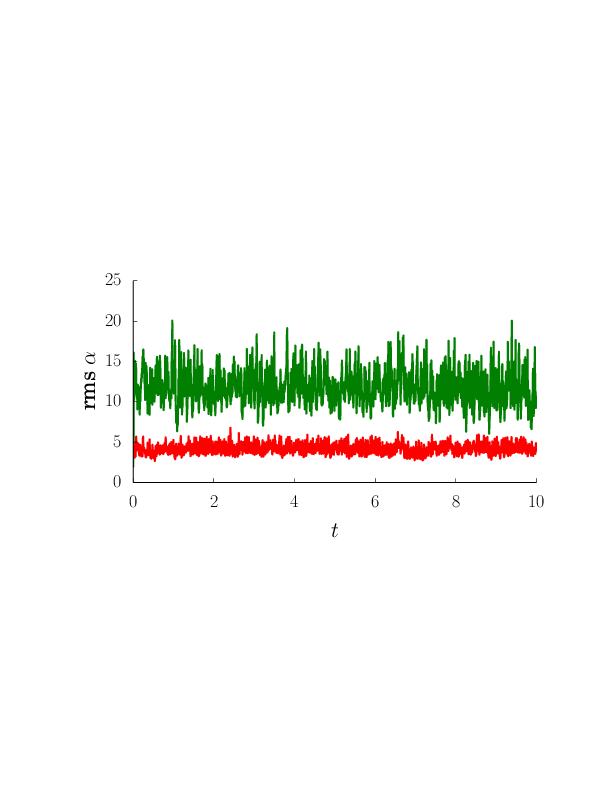}
\vskip -7.2 cm
\hskip 3.5cm \includegraphics[width=\ss]{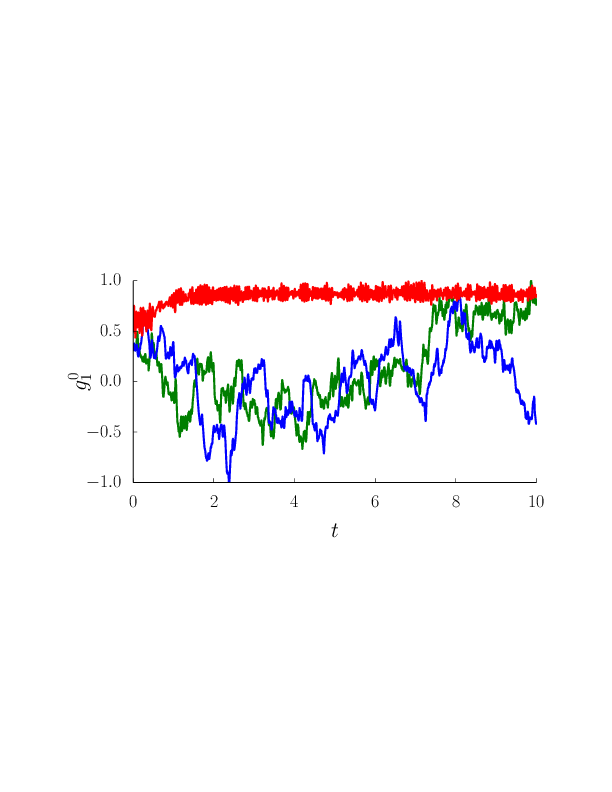}
\vskip -7.2 cm
\hskip 3.5cm \includegraphics[width=\ss]{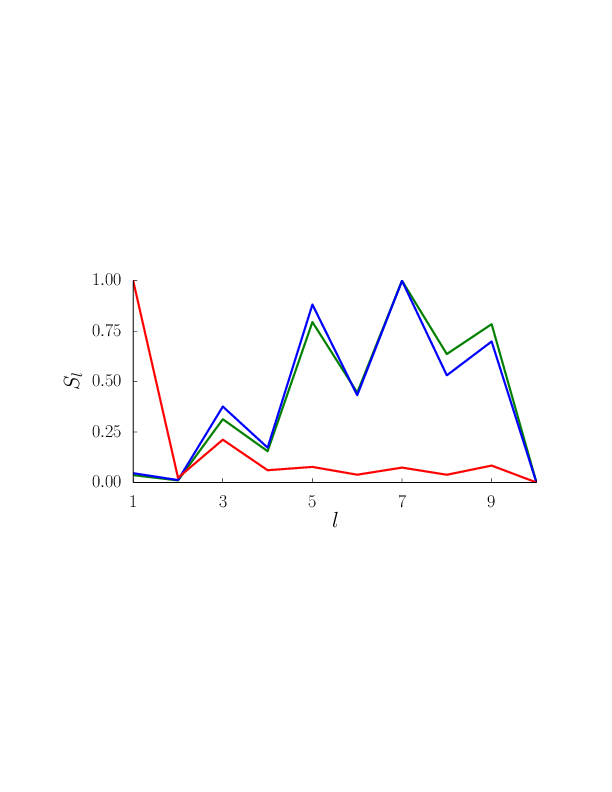}
\vskip -3.5cm
 \caption{Evolution of the root mean square value of $\alpha$ (upper plane), 
  normalized magnetic dipole $g_1^0$   
 (middle plane), and the averaged magnetic spectra $S_l$ (lower plane) for the three regimes: 
    $C_\epsilon= 1$ (red), 7 (green), and 50 (blue).
 The corresponding amplitudes for $g_1^0$:  
 0.31, 0.32, 1.27, 
  and for
 $S_l$: $1\,500$, $9\, 900$, $124\,000$.
} \label{fig3}
\end{figure}


Influence of the $\alpha$ fluctuations on the solution of Eqs(\ref{3}--\ref{4}) was tested, using  our finite difference model, which  can reproduce continuous spectrum up to the scale of the energy injection. The scale of injection is assumed to be the  grid scale, i.e. the  distance between the mesh grid points.

The  random fluctuations modify $\alpha_\circ$ in the  following way: $\alpha_\circ \to \alpha_\circ(1+C_\epsilon \epsilon (r,\,\theta))$, where $\epsilon$ is the uniformly distributed  random variable from -1 to 1, and $C_\epsilon$ is the constant.
 In every mesh grid point $\epsilon$ changed after the time $\delta t=0.01$ simultaneously, see evolution of the 
  root mean square value of $\alpha$ in  
Fig.~\ref{fig3}.
 
We indeed observed appearance of some reversals of the magnetic field, see evolution of $g_1^0$ in Fig.~\ref{fig3}, for $C_\epsilon=7$, 50, which can be related to the geomagnetic field reversals. However, this statement appears to be wrong, because the structure of the magnetic field  spectrum $S_l$ due to fluctuations changed essentially, see Fig.~\ref{fig3}. Before it dissipates at the diffusion scale, the magnetic energy of fluctuations  accumulates at  the wave numbers $l>4$, that is resulted in the appearance of the spectrum's plateau   at $5\le l \le 9$. In other words, to change evolution of the magnetic dipole $g_1^0$ one needs to increase the magnetic energy at the small scales in some orders, see the normalized factors for $g_1^0$ in the figure caption. Such a catastrophic event is hardly believed to happen  in the liquid core if the geomagnetic reversals is treated like the  trivial redistribution of the energy between the harmonics in the white spectrum \cite{R13}.

\section{Geostrophic Regimes}
\label{section:5}
  The specific feature of the planetary dynamo is the geo\-strophic balance of the forces in the liquid core 
 \cite{Pedl}. Assuming that in the leading order viscous and Archimedean forces are small,  one has balance of the Coriolis force and the   gradient of the pressure. Application of the  Taylor-Praudman theorem    leads immediately  to conclusion  that velocity field $\bf V$ is elongated along the axis of rotation. In the other words $\bf V$  in the bulk of the core  depends weakly on the $z$-coordinate. 

In the general case, in presence of the viscous force and the thermal buoyancy,  locations of the large gradients in the $z$-direction correspond to the boundary layers and the equator plane, where physical fields can change the sign.
 This statement relates not only to the large-scale velocity field but to the averaged products of the turbulence, like the kinetic helicity $\chi$, $\alpha$, as well.

\begin{figure}[ht!]
\def\ss{.55}
\def\ssize{9.2cm}
\vskip -2.0cm
\hskip -0.5cm
\begin{minipage}[t]{\ss\linewidth}
\hskip 1.2cm \includegraphics[width=\ssize]{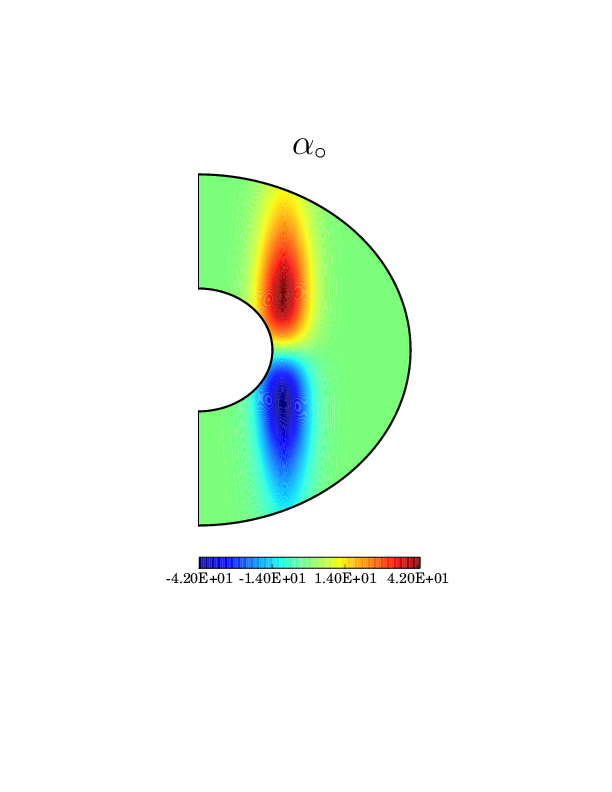}
\end{minipage}\hfill
\begin{minipage}[t]{\ss\linewidth}
\hskip -1.75cm \includegraphics[width=\ssize]{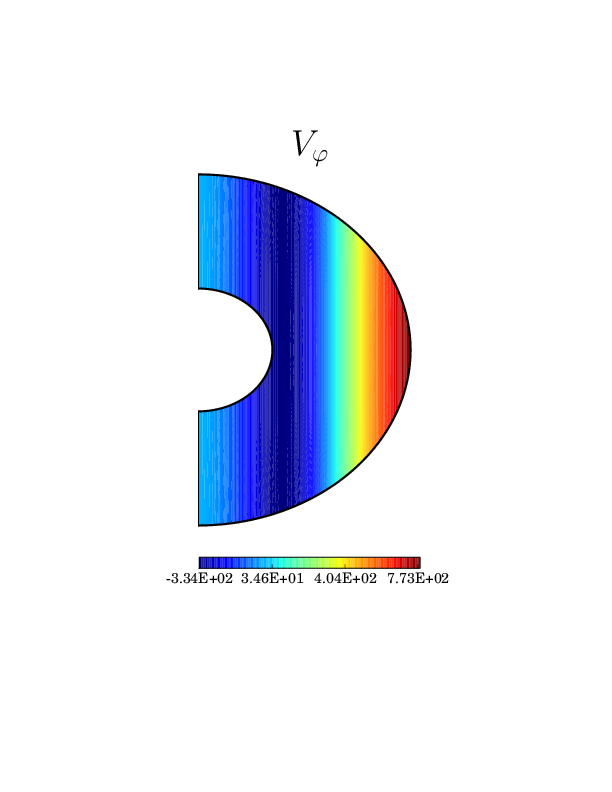}
\end{minipage}
\vskip -2.5cm
 \caption{Spatial distribution of 
 $\alpha_\circ$ and $V_\varphi$ in  the geostrophic regime. 
} \label{fig4}
\end{figure}

Here we use results of 3D simulations of the thermal convection heated from below in the rapidly rotating spherical shell. Roughly,  for the moderate Rayleigh numbers (regime R2 in  \cite{R10}) $\alpha$-effect and azimuthal velocity $V_\varphi$ can be approximated as follows:
\begin{equation}\label{9}
\begin{array}{l}
\displaystyle
\alpha_\circ=C_\alpha \,  r (-erf(1.25 |z|) + 1) e^{\displaystyle-66.7 (s - 0.39)^2}\sin(2\theta)
\\  \\ 
\displaystyle
V_\varphi= C_\omega s \left(e^{\displaystyle-11.76 (s - 0.35)^2} + 0.73 e^{\displaystyle-3.84 (s - 1)^2} \right),
\end{array}
\end{equation}
with the polar coordinates  $s=r\sin\theta$, $z=r\cos\theta$. This approximation corresponds to the convection mainly outside of the Taylor cylinder, see 
Fig.~\ref{fig4}.

The maximum  of $|\alpha_\circ|$ locates near $s=0.45$ and maximum of the radial gradient of the differential rotation, $\displaystyle {\partial \over \partial r}  {V_\varphi\over s}$, is near    $s\sim 0.6$, close to $|\alpha|$'s maximum. It means  that the both sources of generation, $\alpha$-effect, and $\omega$-effect have the same locations, and  meridional circulation will not change solution too much. 

 As we  can expect from Fig.~\ref{fig4}, the scale of the magnetic field in $s$-coordinate 
  is expected to be quite small, because the scales of $\alpha$-effect, and $\omega$ are small as well. This prediction is proved with the simulations
 for $C_\alpha=2980$ and $C_\omega= 35.4$, which are near the threshold of generation. The  ratio of the poloidal to the toroidal energies is equal to 8, and the  maximum of the magnetic energy is at $l=3$. Increase of $C_\alpha,\, C_\omega$ leads to the shift of the maximum of the spectra to the small-scaled part of the spectra. The switch on of the meridional circulation does not help to increase the dipole component of the field.

Magnetic energy oscillates with amplitude about 1\% of its mean value, and amplitude of the magnetic dipole oscillations is even smaller. In spite of the fact that we used $\alpha$ and $V_\varphi$ from 3D simulations,  production of the dipole magnetic field is less efficient than in the model, discussed in the Section 3. 

\section{Dynamic $\alpha$-quenching}
\label{section:6}
The more sophisticated model of $\alpha$-quenching is the so-called dynamic quenching, where the damping of  $\alpha$, given by  the sum $\alpha=\alpha_\circ+\alpha_m$,  is provided with a magnetic part $\alpha_m$, described by the evolution equation \cite{KRR}:
\begin{equation} \label{10}\displaystyle
{\partial  \alpha_m \over \partial t}= 
{\bf B}\cdot {\nabla\times {\bf B}}-\alpha\,
 {{\bf B}^2\over \eta}-{\alpha_m\over {\cal T}}, \end{equation}
where ${\cal T}=1$ is the typical time scale.
The generated $\alpha_m$ has the opposite sign to $\alpha_\circ$ that reduces the total $\alpha$-effect in (\ref{3}).

We tested regimes with $\alpha_\circ$ and $V_\varphi$, given by (\ref{7}--\ref{8}),  and 
  set of parameters close to $C_\alpha=-0.004$, $C_\omega= 30$. The amplitude of the poloidal magnetic energy (1 200) is order of magnitude smaller than the  toroidal part (20 000).  In spite of the quite large values of the magnetic energies, decrease of   $C_\alpha$, and $C_\omega$  at 20-30\% leads to decay of the solution. It can be explained as with the rapid increase of the growth rate in the linearised equations, as well as with coexistence  of two  finite-amplitude solution branches   with the weak and strong magnetic field intensity. Some simulations demonstrate spontaneous transitions from the weak field dynamo to the strong field, accompanied with the reconstruction  of the magnetic energy spectrum that tells in favour of the latter assumption. Such a rapid increase of the  magnetic field production makes it difficult to find a solution with a predominant dipole contribution.

The magnetic dipole $g_1^0$ oscillates, changing its sign with the period $t_{osc}=0.036$.  The magnetic field spectrum has maximum at $l=9$, that corresponds to the small-scaled polarwards dynamo-wave. The further  increase of $C_\alpha$, and $C_\omega$ preserves the zero  mean level of  $g_1^0$.
 This kind of $\alpha$-quenching requires a thorough analysis of the range of parameters, which  can be used for the geodynamo  applications. 

\section{Conclusions}
\label{section:7}
It is quiet expected that the considered above mean-field dynamo do can reproduce some features of the geomagnetic field. At least in principal, $\alpha\omega$-models can generate the predominant dipole magnetic field, similar to that one at the Earth's liquid core, and even the reversals of the field. To the moment it is not clear if the reversal is the intrinsic feature of the dynamo mechanism either it is  triggered with the external perturbation. The both scenarios have its own arguments. Here we showed that even the simple idea of the fluctuating mean-field coefficient, say $\alpha$-effect, should be considered very carefully. Influence of fluctuations on the magnetic dipole evolution should not treated separately from the properties of the magnetic energy spectrum, which can be modified by the fluctuations essentially. 

The other point is the application of the 3D dynamo simulations for estimates of the $\alpha$-effect and differential rotation. Our study reveals that it can not be done straightforward. There are many reasons to that conclusion. One of the reason is that calculation of the averaged quantities like kinetic helicity and $\alpha$ requires the intermediate physical scale, $l_i$, such that  $l_d\ll l_i\ll L$, where
 $l_d$ is the dissipative scale, and $L$ is the scale of the liquid core.
 This is quite difficult task for the 3D simulations, which have still pure resolution. Note also that separation of the scales, well adopted in the  astrophysical applications, is questionable  point in the geodynamo, where the magnetic spectrum is smooth and continuous, and the  intermediate scale can absent at all. We also should not exclude  possibility that some more successful combination of parameters will improve the situation. This problems requires exploration of the phase space and it is a challenge for the cluster computer systems. It will be the next step of the  research in the close future.


\begin{thebibliography}{99}

\bibitem{Bel}
{\it Belvedere, G., Kuzanyan, K.,   Sokoloff, D. D.}
A two-dimensional asymptotic solution for a dynamo wave in the light of the solar internal rotation.
Mon. Not. R. Astron. Soc. 2000. 315. 778--790. 


\bibitem{Br75}
{\it Braginsky, S. I.}
{Nearly axially symmetric model of the hydromagnetic dynamo of the Earth, I}.
 Geomagnetism and Aeronomy. 1975. 15. 122--128. (English translation)

\bibitem{BS05}
{\it Brandenburg, A. and Subramanian, K.}
Astrophysical magnetic fields and nonlinear dynamo theory.
Phys. Rep. 2005. 417.  1--209.



\bibitem{C99}
    {\it Christensen, U. R.,  Olson, P.,   Glatzmaier, G. A.} 
  {Numerical
modelling of the geodynamo: a systematic parameter study}.
{\it     {Geophys. J. Int.} 1999.     {138}},
 {393}--409. 


\bibitem{IGRF}
    {\it Finlay, C. C., 
 Maus, S.,   Beggan, C.D.,   Bondar, N. N.,   Chambodut, A.,   Chernova, N. A.,   Chulliat, A., 
   Golovkov, V. P.,   Hamilton, B.,   Hamoudi, M.,   Holme, R.,   Hulot, G.,    Kuang, W., 
  Langlais, B.,    Lesur, V.,   Lowes,  F. J.,    Luhr, H.,    Macmillan, S., 
   Mandea, M.,    McLean, S.,    Manoj, C.,    Menvielle, M.,    Michaelis, I., 
  Olsen, N.,    Rauberg, J.,    Rother, M.,   Sabaka,  T. J.   Tangborn, A.,    Toffner-Clausen, L., 
   Thebault, E.,   Thomson,  A. W. P..,   Wardinski, I.,   Wei, Z.,    Zvereva, T. I.}
  {International Geomagnetic Reference Field: the eleventh generation}. 
{\it     {Geophys. J. Int.} 2010.      {183}}.
    {3}.      
 {1216}--1230.


\bibitem{HR10}
 {\it    {Hejda}, P. and  Reshetnyak, M.}
  {Nonlinearity in dynamo}.
    { Geophys. Astrophys. Fluid Dynam.} 2010.    {104}.
    {6}.      
 {25}--34.

\bibitem{Hoyng}
    {\it Hoyng, P. }
  {Helicity fluctuations in mean field theory: an explanation for the variability of the solar cycle?}
    { Astron. Astrophys.} 1993.    {272}. 
 {321}--339.

\bibitem{Jones}
    {\it Jones, C. A.}
  {Convection-driven geodynamo models}. 
     {Phil. Trans.~R.~Soc.~London}. 2000.     {A358}.
 {873}--897.



\bibitem{Jouve}
    {\it Jouve, L.,   Brun, A. S..,   Arlt, R.,   Brandenburg, A.,   Dikpati, M.,   Bonanno, A.,   K\:{a}pyl\:{a}, P. J., 
 Moss, D.,   Rempel, M.,   Gilman, P.,   Korpi, M. J.,   Kosovichev,  A. G.}
  {A solar mean field dynamo benchmark}.
     {Astron. Astrophys.} 2008.     {483}. 
 {949}--960.




\bibitem{KRR}
    {\it Kleeorin, N.,  Rogachevskii, I.,   Ruzmaikin, A.}
  {Magnitude of dynamo-generated magnetic field in solar-type convective zones. 
     {Astronomy and Astrophysics}.  1995.    {297}.
 {159}--167.


\bibitem{KM80}
    {\it Kraichnan, R., H.,    Montgomery, D.}
  {Two-dimensional turbulence}.
    { Rep. Prog. Phys.}  1980.   {43}. 
 {547}--619. 


\bibitem{KR}
    {\it Krause,  F.} and  R\"{a}dler, K. H.} 
{Mean-field magnetohydrodynamics and dynamo theory}.
Akademie-Verlag. 1980.

\bibitem{Langel}
    {\it Langel,  R. A.}
  {The main field 
 In Geomagnetism. Ed. J. A. Jacobs}.    {1}.  {249}--512. 
Academic Press.  1987. 


\bibitem{Moss11}
    {\it Moss, D,    Sokoloff, D.,   Lanza, A.F.}
  {Polar branches of stellar activity waves: dynamo models and observations}.
    {Astronomy and Astrophysics}. 2011.     {531}. 
 {A43}.


\bibitem{Moss}
    {\it Moss, D,  Kitchatinov, L. L.,   Sokoloff, D. D.}
  {Reversals of the solar dipole}.
    {Astronomy and Astrophysics}. 2013.     {550}. 
 {L9}.


\bibitem{Pedl}
    {\it Pedlosky, J. }
{Geophysical Fluid Dynamics, 2nd ed.}
Sprin\-ger-Verlag. 1987.

\bibitem{PFL}
    {\it Pouquet, A., Frisch, U.,  Leorat, J.}
  {Strong MHD helical turbulence and the nonlinear dynamo effect}.
    {J. Fluid Mech.} 1976.     {77}.
 {321}--354.

\bibitem{RS03}
    {\it Reshetnyak, M., and  Sokoloff, D.}
  {Geomagnetic field intensity and suppression
of helicity in the geodynamo}. 
    {Izvestiya, Physics of the Solid Earth}. 2003.     {39}.
    {9}.      
 {774}--777.

\bibitem{R10}
    {\it Reshetnyak, M.}
  {Taylor cylinder and convection in a spherical shell}.
    {Geomagnetism and Aeronomy}. 2010.    {50}.
    {2}.    
 {263}--273.

\bibitem{R13}
    {\it Reshetnyak, M. Yu.}
  {Geostrophic balance and reversals of the geomagnetic field}.
    {Russ. J. Earth Sci.} 2013.    {13}. 
ES1001. 

\bibitem{RK13}
    {\it Roberts, P. H.  and King,  E. M.}
  {On the genesis of the Earth's magnetism}. 
     {Rep. Prog. Phys.}  2013.   {76}.
    {9}.      
 {096801}.

\bibitem{Sobko}
    {\it Sobko, G. S., 
 Zadkov, V. N.,   Sokoloff, D. D.,   Trukhin,  V. I.}
  Geomagnetic reversals in a simple geodynamo model. 
    {Geomagnetism and Aeronomy}. 2012.     {52}.
    {2}.      
 {254}--260.

\bibitem{Wicht}
    {\it Wicht, J.}
  {Inner-core conductivity in numerical dynamo simulations}.
    {Phys. Earth Planet. Int.}  2002.   {132}.
 {281}--302.

\bibitem{ZRS}
    {\it Zeldovich,  Ya. B.,   Ruzmaikin, A. A.,   Sokoloff, D. D.}
{Magnetic Fields in Astrophysics}. 
Gordon and Breach Science Pub. 1990. 

\end{thebibliography}
\end{document}